\documentclass[superscriptaddress,reprint,aps,pre,footinbib]{revtex4-1}
\usepackage[english]{babel}
\usepackage[T1]{fontenc}
\usepackage{lmodern}

\usepackage{hyperref}
\usepackage{amsmath}%
\usepackage{amsfonts}%
\usepackage{amssymb}%
\usepackage[usenames,dvipsnames]{color}%
\usepackage{graphicx,tikz}

\newcommand{\addtext}[1]{#1}
\newcommand{\deltext}[1]{}
\newcommand{\explaindeladd}{}

\newcommand{\nubar}{\bar{\nu}}

\newcommand{\siword}{Appendix}
\newcommand{\siabbr}{Appendix}

\makeatletter
\newcommand*{\balancecolsandclearpage}{%
  \close@column@grid
  \cleardoublepage
  \twocolumngrid
}
\makeatother

\begin{document}
\title{Universality and scaling laws in the cascading failure model with healing}
\pacs{89.75.Hc, 64.60.ah, 05.10.-a}
\author{Marcell Stippinger}
\email{stippinger@phy.bme.hu}
\affiliation{Department of Theoretical Physics, 
Budapest University of Technology and Economics, 
H-1111 Budafoki \'ut 8., Budapest, Hungary}
\author{J\'anos Kert\'esz}
\email{KerteszJ@ceu.edu}
\affiliation{Center for Network Science,
Central European University,
H-1051 N\'ador u. 9., Budapest, Hungary}
\affiliation{Department of Theoretical Physics, 
Budapest University of Technology and Economics, 
H-1111 Budafoki \'ut 8., Budapest, Hungary}

\explaindeladd

\begin{abstract}
Cascading failures may lead to dramatic collapse in interdependent networks, where the breakdown takes place as a discontinuity of the order parameter.
\addtext{In the cascading failure (CF) model with healing there is a control parameter which at some value suppresses the discontinuity of the order parameter. However, up to this value of the healing parameter} the breakdown is a hybrid transition, meaning that, besides this first order character, the transition shows scaling too.
\addtext{In this paper we investigate the question of universality related to the scaling behavior.} Recently we showed that the hybrid phase transition in the original CF model has two sets of exponents describing respectively the order parameter and the cascade statistics, which are connected by a scaling law. \addtext{In the CF model with healing we measure these exponents as a function of the healing parameter.} We find two universality classes: \addtext{In the wide range} below the critical healing value the exponents agree with those of the original model, while above this value the model displays trivial scaling \addtext{meaning that fluctuations follow the central limit theorem}.
\end{abstract}
\maketitle

\section{Introduction}
\label{sec:intro}

Coupled infrastructural networks are extremely vulnerable to cascading failures~\cite{Buldyrev2010}. Buldyrev \emph{et al.}\ introduced the concept of interdependent networks~\cite{Buldyrev2010} in order to elucidate the mechanism behind this observation. \addtext{This cascading failure (CF) model consists of two layers of networks, and, in addition to the intra-layer connectivity links, it introduces so-called dependency links that model the need for resources or services coming from the other the layer.} 
This model exhibits rich behavior, among others it shows a hybrid phase transition (HPT), where the order parameter has both a jump and critical scaling.
The CF model has been extended in several ways, e.g. dependency links were limited to finite range~\cite{Li2012} to capture the cost one must pay for long-range connections, multiple layers were considered~\cite{Havlin2012,Buldyrev2012} to model a variety of interconnected infrastructures and partial dependence~\cite{Zhou2013} allowed some nodes to be autonomous on some resources. \addtext{These earlier works focused only on the necessary conditions for the breakdown but now the repairing of interdependent networks is getting into focus too.}

\addtext{Some repairing strategies involve the random recovery of a portion of the existing infrastructures. This has been modelled in a probabilistic cellular automaton for interdependent networks~\cite{Majdandzic2016} and in the original CF model~\cite{DiMuro2016}.
These strategies assume that the original components and links are repairable and they act respectively on the two layers in a mean-field way and at the mutual boundary of the still functional component.
Crisis situations, in contrast, where the original components cannot be recovered, are better described by the dynamic reorganization of existing components.} A \deltext{dynamical} \addtext{stochastic} healing rule was introduced~\cite{Stippinger2014} that describes the efforts spent on repairing the network via new links on longer timescales.

Although the original model exhibits HPT, it was shown that control parameters, such as the range of dependency links or the healing probability, allows for eliminating the jump in the order parameter, i.e., changing the transition from hybrid to second order~\cite{Li2012,Stippinger2014}.

Recently it was found that the hybrid transition can be characterized by two sets of exponents, one for order parameter and another one for the statistics of finite cascades~\cite{Lee2016}. These are related by a scaling law, which connects the exponent of the order parameter to that of the first moment of the cascade size distribution. Calculations were carried out on the square lattice and the Erd\H{o}s--R\'enyi network (corresponding to the mean field case). A very efficient algorithm is needed to calculate these critical parameters~\cite{Hwang2015}. This is probably the reason, why little effort has been devoted to the problem of universality in such systems. 

In this paper we study an extension of the original CF model. First, we extend the simulation algorithm to interdependent networks with healing. Next, we identify two universality classes separated the critical healing probability: One bearing a hybrid phase transition and a continuous one described by trivial scaling exponents. Finally, we argue that networks close to the critical healing probability are mixtures of network realizations from below and above the critical healing therefore their behavior is ambiguous and not well characterized by scaling exponents.

\section{The cascading failure model}
\label{sec:healingmodel}

Interdependent networks mutually supply and depend on each other. This is captured by
the cascading failure (CF) model~\cite{Buldyrev2010} that is built of two topologically identical starting networks $A$ and $B$\deltext{ with usual connectivity links}. \addtext{These networks are basically graphs: both of them have the same number of nodes $N$ and the usual edges are termed connectivity links. Here we consider special graphs organized in periodic structure forming square lattice networks with nearest-neighbor connectivity links within each layer.}

The relationship between the layers is expressed by random dependency links producing a one-to-one mapping of the nodes of the two layers. Dependency link means that if a node fails, its dependent fails too and \deltext{is} \addtext{their links are} removed.
Only mutually connected components (MCC) are viable that is nodes belonging to a MCC must be connected via connectivity links in layer $A$ and their dependents in layer $B$. \addtext{The nodes of each MCC form a subset of a usual connected component in layer $A$ and, similarly, in layer $B$ too.}
Due to this restriction, the failure of a single node may result in fragmentation of connected components in a layer that triggers splits \addtext{in the other layer, and so on, the failure might propagate back and forth between the layers. This iterative process resulting in the MCCs is} referred to as failures cascading between the layers.

The robustness of the network is studied, as follows. At the beginning one layer is subject to a random attack in which failure is induced externally in $1-p$ fraction of the original nodes. \addtext{Then the mutually connected components fulfilling the above restrictions are determined.} Finally, the size $m$ of the remaining giant (largest) mutually connected component is measured relative to the original network. \addtext{The iterative steps of determining the mutually connected components can be viewed as a critical branching process as described in~\cite{Zhou2014}. Although this is a simple dynamic interpretation of the original model, in the next section we define a more complex dynamics which takes into account the whole history of the network that tries to adapt and reorganize itself due to the external failures.}

\subsection{CF model with healing}
\label{sec:healing-algo-original}
The cascading failure model with healing~\cite{Stippinger2014} is a dynamic version of the CF model \addtext{in the sense that there is an external source of failures which targets the nodes one-by-one. This is a one-way process and} the nodes of network layer $A$ are targeted gradually, in a random order. In each step the next node in the ordering is ``targeted'' and a successful ``attack'' is carried out if the node is part of the giant mutually connected component. (The fraction of targeted nodes is $1-q$ while the fraction of attacked nodes is $1-p$, it follows from the definition that $1-q>1-p$, therefore $p>q$. \addtext{The difference between $p$ and $q$ is $\int_q^1 \big((\tilde{q}-m(\tilde{q}))/\tilde{q}\big)\,\mathrm{d}\tilde{q}$ as derived in~\cite{Stippinger2014}.}) The \addtext{unavoidable} failure of the targeted node is induced by removing all of its connectivity links. After that the following dynamics is applied to relax the network \addtext{and form the new mutually connected components with healing links added}:
\deltext{
\begin{enumerate}
\item Make a list of the nodes that lost any links. For each node in the list take each pair of its previous neighbors that survived the removal and connect them with independent probability $w$ if there is no link between them and their dependents are in a connected component in the other layer.
\item Then, in the other network, remove those links of dependent counterparts that run between nodes that are no longer connected.
\end{enumerate}
}
\begin{enumerate}
\addtext{
\item \label{alg-heal-depend} Remove the links running between nodes whose dependent counterparts are no longer in the same connected component.
\item \label{alg-heal-link} Make a list of the nodes that lost any links in since the previous iteration. For each node in the list take each pair of its previous neighbors and propose them as a healing link with independent probability $w$. Realize the proposed links if they do not exist and their dependents are in the same connected component in the other layer.
}
\item \label{alg-heal-repeat} Repeat \ref{alg-heal-depend} to \ref{alg-heal-link} on all layers until no more links are removed.
\end{enumerate}
The creation of new links in Step~\ref{alg-heal-link} can be interpreted as the effort to find new partners which replace the failed ones. \addtext{These healing links change the original topology of the network~\cite{Stippinger2014}. Note that this is an extended version of the original healing algorithm to handle multiple connected components and we show that in the infinite system limit this modification has no significant impact on the measured quantities.}
After the network is relaxed, one proceeds with the next step in which the next node on the list is targeted. The procedure is pursued until the connectivity links cease to exist. The principal quantity of interest is the order parameter, which is the relative size $m$ of the giant mutually connected component as compared to the original size of the network. For small healing probability $w<w_c\approx 0.355$ the model exhibits a first-order (hybrid) phase transition while for $w>w_c$ the phase transition is of second-order \cite{Stippinger2014}.

\section{Simulation method} 
\label{sec:algorithm}
Phase transitions are often accompanied by scaling laws but the numerical test of the relevant quantities for the CF model had been a challenging task. Recently, however, efficient algorithms have been developed, that allow for large scale simulation of the mutually connected components (MCC) in the CF model with computational time of $O(N^{1.2})$ \cite{Hwang2015,Grassberger2015}.

The implementation of \cite{Hwang2015} is based on the idea that connectivity links only within MCCs are kept and the other ones are \emph{inactivated}. Consider a node removal (and the removal of its links as a consequence) in the layer $A$ that splits up a component. Then the new MCCs are to be found. The split is propagated to layer $B$ by inactivating all the links that bridged the newly split components in layer $B$. Of course this might trigger further splits 
that must be propagated back to layer $A$ and so on. This algorithm becomes very efficient using a proper graph data structure.

The underlying fully dynamic graph algorithm (see \cite{Holm2001}) can account for connectivity in a single layer and this accounting is efficient both for adding and removing edges. On each edge removal or inactivation it detects if the component in the layer is split. Then one can query the size of the new components and optionally the members of any of these or even whether two nodes belong to the same component, all of this very efficiently.

The application of algorithm \cite{Hwang2015} to the healing problem needs some effort.
Notably, the algorithm must integrate the step of adding healing links efficiently which is not obvious.
In the following we generalize the algorithm to efficiently simulate interdependent networks with healing.

In the cascading failure model only the \deltext{largest} \addtext{giant} mutually connected component (GMCC) is of interest. By deleting links the GMCC and the smaller mutually connected components get fragmented into smaller components. However, adding healing links might save a component from the fragmentation. Therefore the deletions are not to be propagated immediately.

While the split of a component is easily propagated, the inverse, notably adding a link between two components in one layer can be computationally expensive. Adding a link requires a search to find which of the previously inactivated links need to be reactivated to find the GMCC. This search would possibly involve many small MCCs.
The design challenge lies in rewriting the steps of the previous healing algorithm in such a way that healing links are added within components only. As a consequence it is assured that manipulation avoid reuniting components, i.e., components are always left intact or split. This choice is justified in the following.

To fulfill the above constraints we propose an algorithm that buffers the links to be deleted while it adds the healing links before actually deleting or inactivating any links in the given layer, see the Algorithm 
in the \siword.
\addtext{We find that the runtime of our algorithm scales as $O(N^{1.3})$ below $w_c$ and at most as $O(N^3)$ near or above the critical point, see~\autoref{fig:heal-runtime-suppl} in the \siword.
This runtime cannot be directly compared to that of the wave algorithm proposed in~\cite{Grassberger2015} because of the following two reasons. First, the runtime $O(N\langle h\rangle)$  of the wave algorithm is given for a single external attack. Second, the number of waves (iterative steps) is known analytically to be $\langle h \rangle\sim O(N^{1/3})$ for ER-networks only~\cite{Zhou2014}. Assuming that in most cases the waves need to be fully recalculated after an external failure, the wave algorithm has at least $O(N^2)$ total runtime for $w=0$ (and exactly $O(N^{7/3})$ for ER-networks). In comparison, our algorithm is intended for fast updates at the cost of higher memory use but for the healing model the runtime is the bottleneck.}

With the generalized algorithm, we can investigate critical properties of the hybrid percolation transition of the CF model with healing thoroughly. We measure various critical exponents including susceptibility and correlation size that were missing in previous studies for the healing-enabled version of the two-dimensional (2D) lattice interdependent networks.

\section{Results on the CF model with healing}
\label{sec:simulation}
In the study of the healing we choose a 2D embedding topology with two $N=L\times L$ square lattices, both with periodic boundary conditions. Each node in one layer has a one-to-one dependent node in the other layer.

The number of externally removed nodes is controlled. We define the control parameter $p$ in the view of the one-by-one removal, \deltext{i.e., in each time step a still functional random node is externally attacked and removed} \addtext{and time is measured in the number of successful attacks. That is, random nodes are targeted externally and they get disconnected by removing their links. The time is increased if the targeted node belonged to the GMCC.} The link removal is eventually followed by a cascade of failures. At the end only the new giant mutually connected component is considered functional. This measurement of time is justified by the fact that smaller MCCs consist of a few nodes only and are often considered not viable~\cite{Li2012}. The fraction of original nodes attacked externally is denoted by $1-p$.

We simulated system sizes \addtext{$N = 32^2$, $64^2$, $128^2$}, $256^2$, $512^2$, $1024^2$, and $2048^2$, with $384$ network configurations for the largest system and increasingly more for smaller systems (\makebox{\# configurations} $\approx 1.6\cdot 10^9/N$).

\subsection{Critical behavior of the GMCC}

The order parameter $m(p)$ is defined as the size of the GMCC per node, which shows the typical behavior of the order parameter at a hybrid phase transition:
\begin{equation}
m(p) = \left\{\begin{array}{ll}
        0, & \text{for } p < p_c \\
        m_0+u\,(p-p_c)^{\beta_m}, & \text{for } p\geq p_c.
        \end{array}\right.
\label{eq:hybrid-transition}
\end{equation}

The critical values $m_0$ and $q_c$ have been published~\cite{Grassberger2015} with great accuracy for the square lattice without healing ($w=0$). \deltext{We used this case as a benchmark test.} \addtext{We measured $p_c$ and $q_c$ too and used this case as a benchmark test.} We calculated the critical values for various $w$ parameter settings using our algorithm. 

We check first whether the simulation results are in agreement with previous findings. The critical exponent $\nubar_m$ is defined via finite size behavior of the interdependent network
\addtext{\footnote{\addtext{This quantity is different from $1/\alpha$ introduced in \cite{Zhou2014} as explained in the following. In the case the avalanche is triggered by an initial removal of a fraction of the nodes the exponent $1/\alpha$ describes the finite size scaling of the avalanche duration (the number of back-and-forth propagation steps). However, in the healing model the avalanches are triggered by repeated single node removals therefore the avalanche dynamics is different. In this case a similar quantity, the exponent $\nubar_a$ was introduced but they may eventually differ in value.}}}
 as $\sigma\propto N^{-1/\nubar_m}$ where $\sigma$ stands for the standard deviation of the critical point $p_c$. (For sake of simplicity, we incorporated the dimensionality $d=2$ into the definition $\nubar_m \equiv \nu_m d = 2\nu_m$.
\addtext{Note that for all data points the smallest number of network realizations was $384$ therefore the bias introduced due to the number of runs is negligible as demonstrated in the following. We used bootstrapping to measure the mean of the absolute difference between \emph{the standard deviation estimated from $384$ realizations} and the population standard deviation (estimated from all available data). This difference was about $3\%$ of the population standard deviation. Over the interval of the seven system sizes this might introduce $0.01$ absolute error in the slope measured on the $\log$--$\log$ plots when the average errors are set methodically to get the largest effect. As known, \emph{the estimation of the variance from realizations} is unbiased if the degrees of freedom is one less than the number of realizations. Therefore this error can be neglected compared to other fluctuations that appear in the numerical values of the exponents.}) In case of $w=0$ the exponent was previously measured $\nubar_m=2.2\pm 0.2$ \cite{Lee2016}.

\begin{figure}
\includegraphics{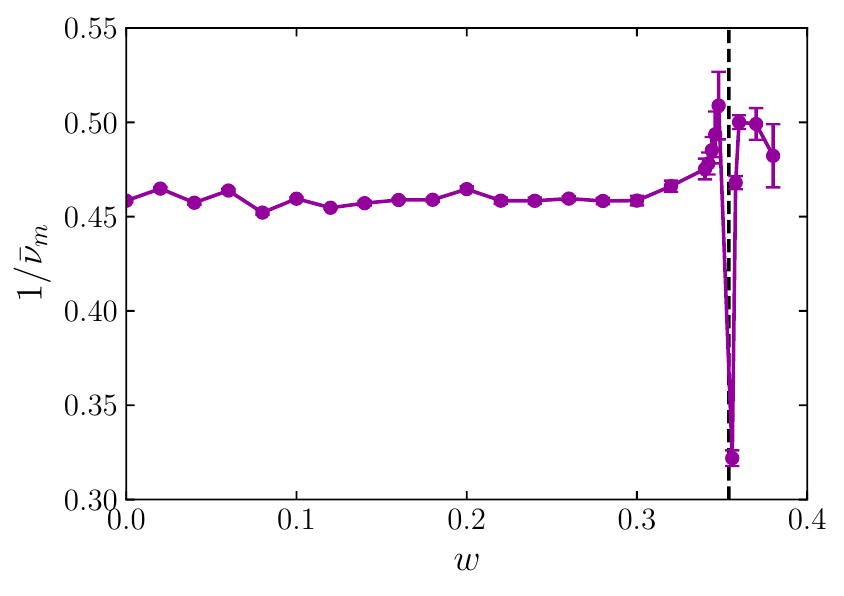}
\caption{(color online) The scaling exponent $\nubar_m$ takes two distinct values: $\nubar_m=2.2\pm 0.1$ for $w<w_c$ and $\nubar_m=2.0 \pm 0.1$ for $w>w_c$ indicating two universality classes separated by the critical value of the healing parameter $w_c$. In the latter regime the system exhibits trivial scaling in the sense that the variance is inversely proportional to the number of nodes in the system. Near $w_c$ the sample variance is dominated by the stochastic mixing of systems that are above and below their perceived critical point. This makes the measurement of $\nubar_m$ unreliable.}
\label{fig:heal-nu-m}
\end{figure}

For the network with healing the above value for $\nubar_m$ persists (see \autoref{fig:heal-nu-m}\addtext{ and \autoref{fig:heal-nu-m-suppl} in the \siword}) until approaching $w_c$.
Above $w_c$ it gets stabilized at $\nubar_m\approx2$ indicative of trivial scaling behavior in the sense that the standard deviation is inversely proportional to the square root of the number of nodes in accordance with the central limit theorem.
Near $w_c$ we could not disclose the true value of $\nubar_m$ because of the following reason. Trying to simulate a system with a specific $w$ near $w_c$ results in an ensemble of systems mixing scaling behaviors below and above $w_c$. This mixing has an extra contribution, in addition to and dominating over the sample variance. The large $\nubar_m$ indicates that the ``unintentional'' mixing part depends less on system size, at least for the system sizes that are accessible even with our efficient algorithm.
The extra variance necessarily makes the distribution of the critical point wider as a consequence, it also makes it rather difficult and unreliable to extrapolate the critical point for the infinite system. In fact, the scaling breaks down and the determination of the critical exponents is hampered by the above effect in this regime.

\begin{figure}
\includegraphics{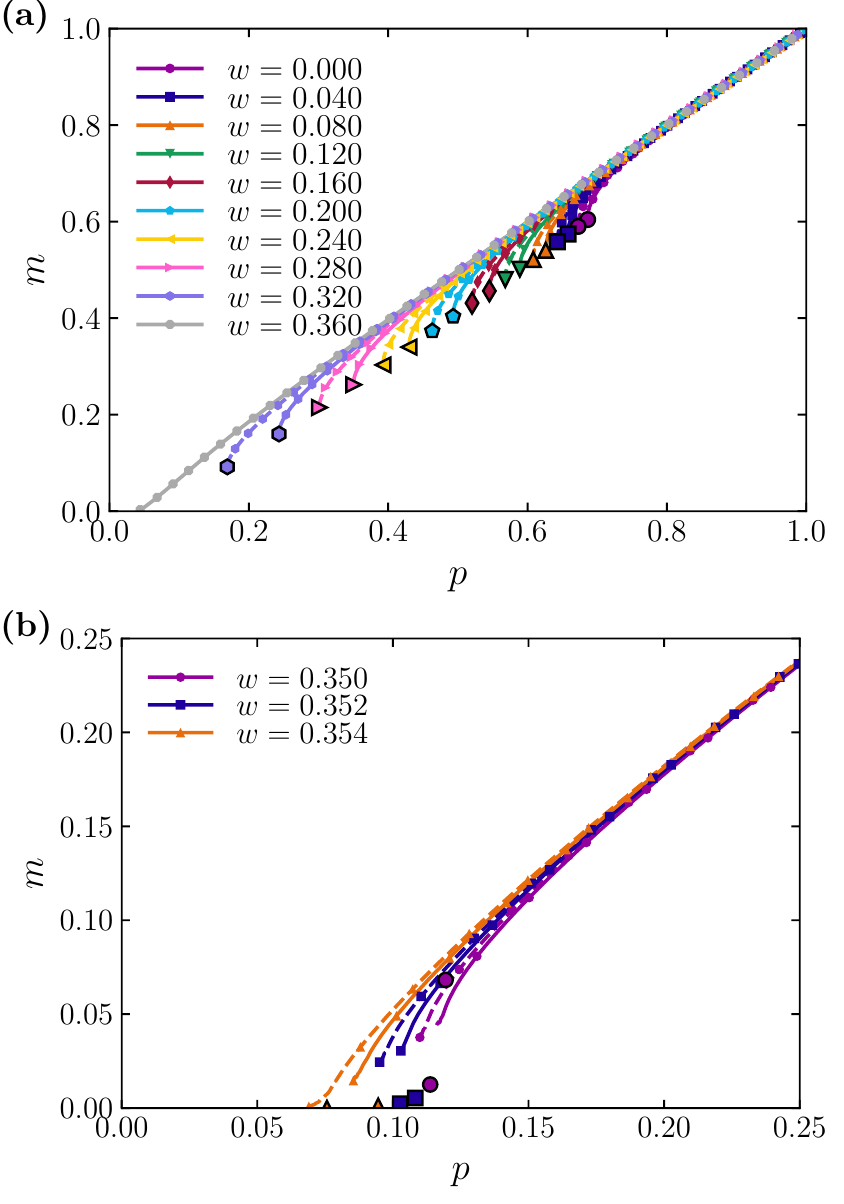}
\caption{(color online) Critical behavior of the order parameter as obtained as an average over systems of size $N=2048^2$. The solid curves are listed from right to left, dashed lines of the same color represent following $w$ in between using equidistant steps.
Large symbols mark the \deltext{breakdown point} \addtext{critical point $p_c$ and the related jump of the order parameter $m_0$}, both of them extrapolated for infinite systems using finite size scaling. (a) An overview, (b) zoom-in to show that for $w\geq0.355$ the phase transition is of second order. }
\label{fig:heal-orderparam}
\end{figure}

The dependence of the order parameter on $p$ and $w$ is shown in \autoref{fig:heal-orderparam}. We define the value of $w_c$ as the smallest $w$ for which we observe a continuous phase transition and we find $w_c=0.355$. This value agrees well with the $w$ where $\nubar_m$ has a sharp maximum. The scaling $1-m(p,w)=a(w)-a(w)\,m\left(1-\frac{1-p}{a(w)}, 0\right)$ that is asymptotically satisfied in the $w\to 0$ limit \cite{Stippinger2014} is confirmed by the new measurements. $a(w)=(1-p_{c}(0)-\Delta p_c(w))/(1-p_{c}(0))$ where $\Delta p_c(w)\equiv p_c(w)-p_c(0) \propto \,w^\delta$ and the exponent value is $\delta=1.006\pm0.009$.

\begin{figure}
\includegraphics{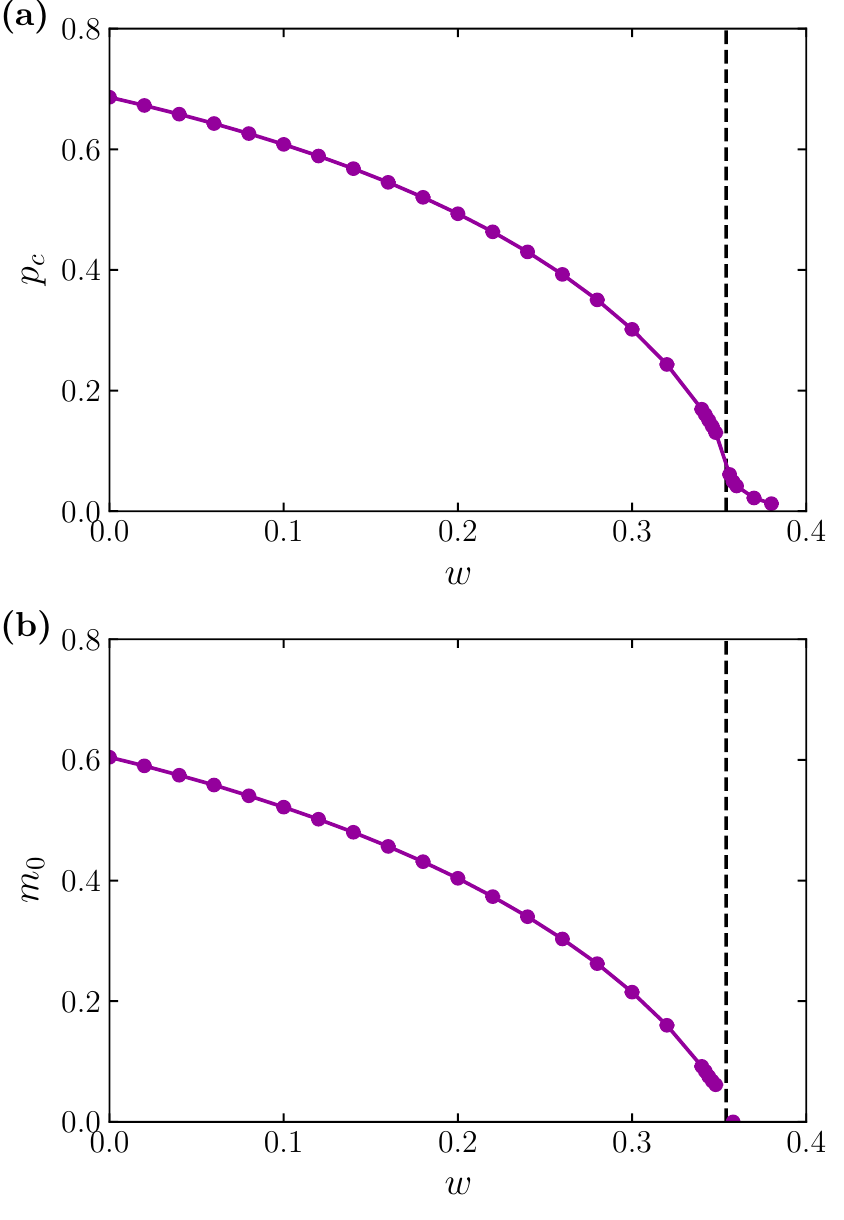}
\caption{(color online) (a) The \deltext{transition} \addtext{critical} point $p_c$ and (b) \deltext{the size of the last avalanche} \addtext{the related jump of the order parameter, i.e., the size of the infinite avalanche} $m_0$. As the healing probability $w$ increases, $m_0$ converges smoothly to $0$ and macroscopic cascades get finally eliminated at $w_c$.
The critical fraction of unattacked nodes $p_c$ is still non-zero at $w_c$ indicating that some small cascades still exist. Only at $w=1$ they cease to exist. }
\label{fig:heal-critical}
\end{figure}

The critical control parameter value $p_c$ and \deltext{the size of the breakdown} \addtext{the related jump $m_0$ of the order parameter} (see \autoref{fig:heal-critical}) are extrapolated by finite size scaling, see \autoref{fig:heal-critical-suppl} in the \siabbr.
For $w>w_c$ the size of the breakdown $m_0=0$ indicating that macroscopic cascades do not occur.
During the process when eliminating nodes one by one small cascades may occur; the smallest cascades cease to exist only at $w=1$, see \autoref{fig:heal-critical}.

\begin{figure}
\includegraphics{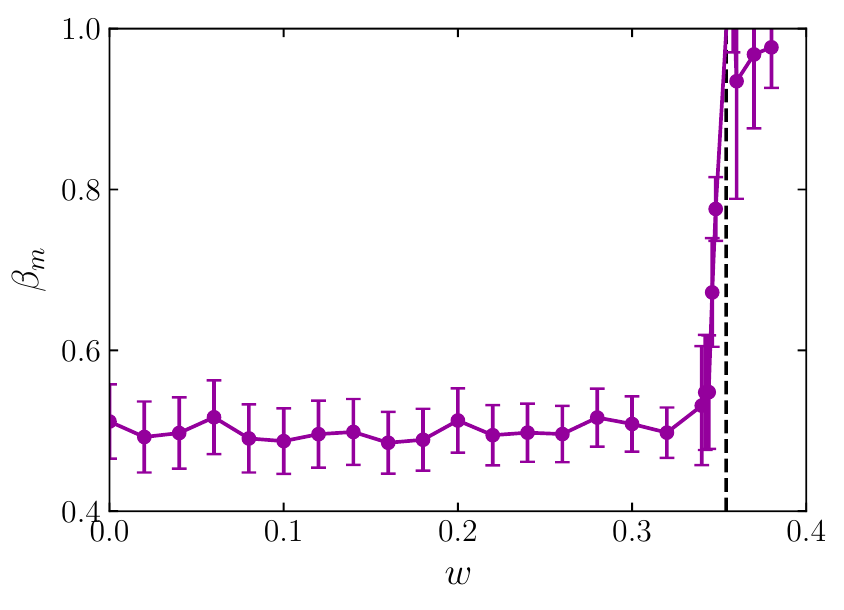}
\caption{(color online) The exponent $\beta_m$ characterizes the change of the order parameter above the jump $m_0$ in the critical regime. $\beta_m = 0.5\pm 0.05$ for $w<w_c$ which confirms the theory for a universality class, and reproduces $\beta_m = 0.5$ for $w=0$, as proved in \cite{Lee2016}. For perfect healing ($w=1$) the value $\beta_m=1$.
}
\label{fig:heal-beta-m}
\end{figure}

The scaling exponent $\beta_m$ quantitatively describes the order parameter in the scaling domain of the hybrid phase transition and is defined according to \eqref{eq:hybrid-transition} as $(m(p)-m_0) \propto (p-p_c)^{\beta_m}$. Without healing, the previously obtained and analytically proved $\beta_m = 0.5$ \cite{Lee2016} is reproduced and it holds up to very close to $w_c$, see \autoref{fig:heal-beta-m}.
This value seems to be universal for $0\leq w<w_c$.

These results suggest that there exist two universality classes. One is characterized by the hybrid transition with $\beta_m=0.5$ and $\nubar_m\approx 2.2$.
The other universality class has a vanishing giant component at breakdown and is characterized by exponents $\nubar_m=2$ and $\beta_m=1$. For the latter, see Figs. \ref{fig:heal-nu-m} and \ref{fig:heal-orderparam}, as well as, \autoref{sec:perfect-healing}.

\begin{figure}
\includegraphics{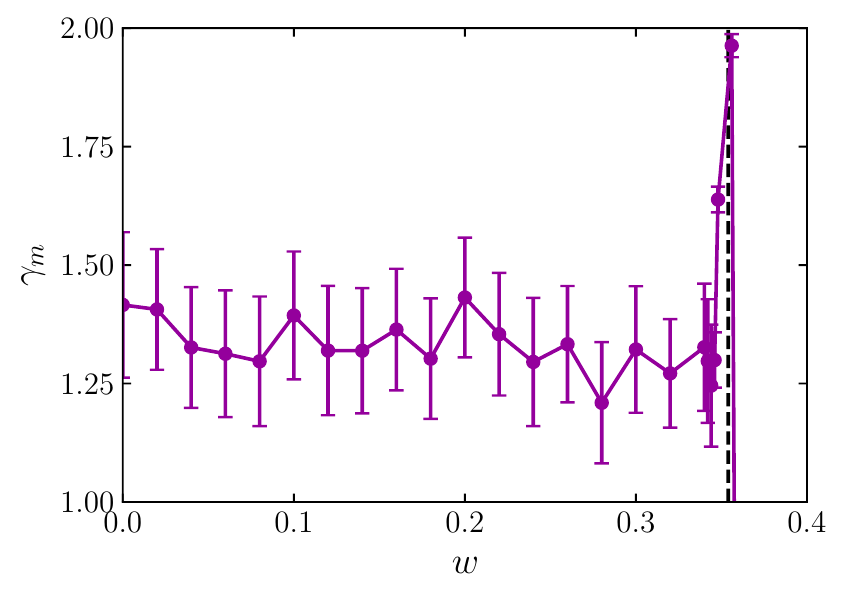}
\caption{(color online) The exponent $\gamma_m$ describes the scaling of the susceptibility close to the critical point. Due to lack of data we are unable to measure $\gamma_m$ in the region $w>w_c$.}
\label{fig:heal-gamma-m}
\end{figure}

The exponent $\gamma_m$ is defined by the scaling of the susceptibility $\chi\equiv N(\langle m^2\rangle-\langle m\rangle^2)\propto (p-p_c)^{-\gamma_m}$. Unfortunately, we were able to calculate the $\gamma_m$ values only with rather large error bars. They scatter between $1.41\pm 0.15$ and $1.33\pm0.15$ (see \autoref{fig:heal-gamma-m}). We conclude that they do not contradict the assumption of universality.
Due to lack of data we are unable to measure $\gamma_m$ in the region $w>w_c$. Later we will present an argument that the exponent $\gamma_m$ should be $0$ in this region.

\subsection{Critical behavior of avalanches}
\label{sec:heal-avalanches}

\begin{figure}
\includegraphics{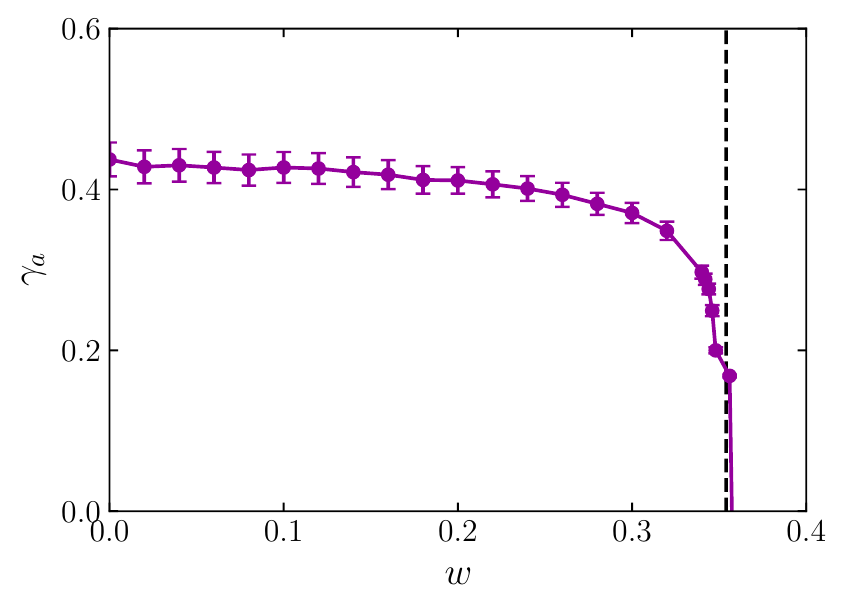}
\caption{(color online) The exponent $\gamma_a$ shows that the average avalanche size increases as the system gets close to the breakdown. Applying healing above $w_c$ stops most avalanches, $\gamma_a=0$. Simulation near $w_c$ we get a mixture of systems above and below the critical healing due to finite size effects therefore our measurement of $\gamma_a$ gets imprecise.}
\label{fig:heal-gamma-a}
\end{figure}

\begin{figure}
\includegraphics{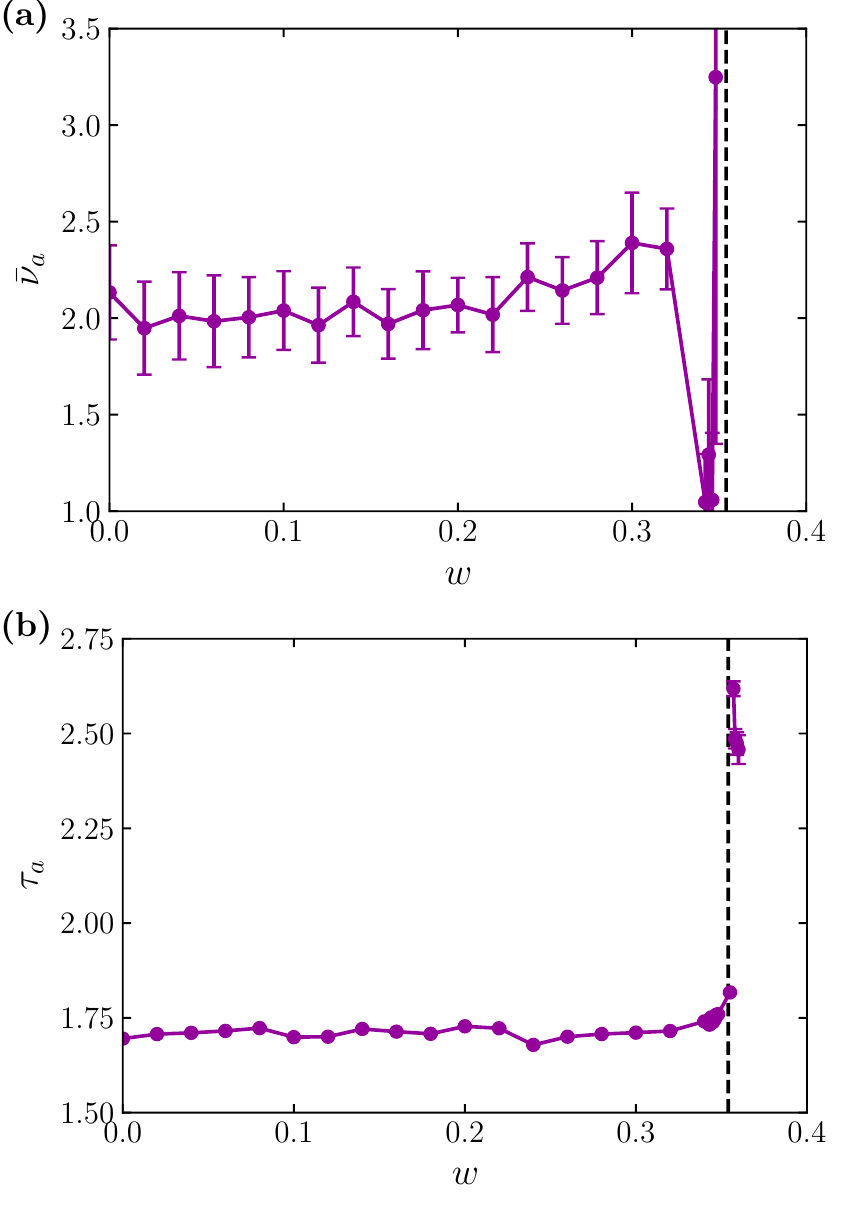}
\caption{(color online) The exponents (a) $\nubar_a$ and (b) $\tau_a$ are approximately constant for $w<w_c$ showing that the universality classes discovered for $\beta_m$ and $\gamma_m$ extend to exponents related to avalanches too. Close to the critical point $w_c$ large cascades get less likely and these exponents cannot be well defined.}
\label{fig:heal-tau-a}\label{fig:heal-nu-a}
\end{figure}

There is another set of critical exponents \cite{Lee2016}, $\tau_a$, $\sigma_a$, $\gamma_a$ and $\nubar_a$ describing the statistics of avalanches. They can be evaluated only for $w<w_c$ where the number of avalanches is sufficient and avalanches are governed by scaling laws.
The size $s$ of the avalanche is the number of nodes failing due to a single external attack\addtext{, in other words, this is the shrinking of the GMCC in one time step}. The finite avalanches are those that happen before the breakdown.
\addtext{One can get a reliable statistics of the finite avalanches by simulating only a reasonable number of network realizations as we did. The idea is that in finite systems close enough to $p_c$ all finite avalanches are in accordance with the critical scaling therefore one can use the avalanches in $[p_c, p_c+\Delta p]$ for $\Delta p < \sigma$. We $\Delta p = 0.25\sigma$ because it yields many avalanches and critical behavior is assured.}

The exponent $\gamma_a$ is defined with the average size of the finite cascades $\langle s_{\text{finite}} \rangle \propto \Delta p^{-\gamma_a}$ for which $1-\beta_m = \gamma_a$ holds \cite{Lee2016}. This relationship is confirmed reasonably well for $w\leq0.1$ where sufficient data is at our disposal, see \autoref{fig:heal-gamma-a}. For $w>w_c$ we would need even larger samples then studied to have sufficient statistics.
Near $w_c$, however, the previously described mixing of realizations of network states from both continuous and discontinuous transitions makes our estimate for $p_c$ less reliable. As a consequence the distance $p-p_c$ from the critical point is less reliable too making it difficult to measure critical exponents.\footnote{Unfortunately one cannot gain insight by experimenting with $p_c$ to get satisfactory scaling because $m_0$ is to be determined in parallel making the experimentation prone to errors.}
The other exponents are defined as $ \left. p_s \right|_{N=\infty} \propto \Delta p^{-(\tau_a-1)/\sigma_a} $ and $ \left. p_s \right|_{\Delta p=0} \propto N^{(\tau_a-1)/\nubar_a\sigma_a} $. The measurements support our hypothesis of a single universality class for avalanche-related exponents below $w_c$, see \autoref{fig:heal-tau-a}. The avalanche-related exponents are meaningless above $w_c$ therefore we do not analyze them.

\subsection{Behavior near $w=1$}
\label{sec:perfect-healing}
We have confirmed numerically that there is a critical value $w_c$ for the healing above which macroscopic cascades disappear and the network tends to get more and more connected. Here we focus on the critical behavior close to $w=1$ and we prove that $\beta=1$ for this case.

In a square lattice 
between any two nodes $U$ and $V$ there exist initially at least two disjoint paths that only have $U$ and $V$ in common. Whenever we externally remove a node (different from $U$ and $V$) at least one of those paths remains intact. As a consequence all the remaining nodes remain attached to the giant component. Thanks to the perfect healing ($w=1$) all possible bridges over the removed node are formed. This way the eventually cut paths between $U$ and $V$ are re-established and again there will be at least two disjoint paths between any two nodes. Correspondingly, in the case of interdependent layers, the removal of one node causes only the removal of its interdependent counterpart and no avalanches are induced:
$m_0=p_c=0$, $m\equiv p$, therefore $\beta_m=1$, $\chi_m\equiv0$ and it does not make any sense to calculate $\gamma_m$ or the exponents related to avalanches.

When $w\lessapprox 1$ avalanches might occur but they are rare and small. For example, to initiate a cascade at least a small region of $n$ nodes one needs to get separated from the connected component in one of the layers. To achieve this at least the perimeter of that region must be cut. In two dimensions the length of the perimeter is at least $\propto \sqrt{n}$. Cutting means that healing links are not allowed to form bridges. This happens with probability at most $(1-w)^{\sqrt{n}}$.
When separated, the propagation of the damage from the initial $n$ nodes may lead to a cascade of size $s\geq n$. As the dependency links have here unlimited range, the counterparts of the original nodes are far away from each other and it has small probability that their failure will lead to further separation of other components because such a separation must be prepared similarly. That is, for separating a single node in the 2D case at least three other connectivity links are needed to be cut previously without healing. This happens with probability smaller than $(1-w)^3$. So the typical cascades are of small size and one iteration. This has been confirmed by simulations.
This means that, when approaching the critical point $p_c\approx0$, the network is so densely connected that cascades are prevented almost surely so $\beta_m=1$ holds also in the vicinity of $w=1$. It is tempting to conclude that this observation points toward the existence of universality for $w > w_c $.
Assuming that the small avalanches are mostly independent their number per unit cell is determined by the central limit theorem, that is the fluctuation of their number is inversely proportional to the square root of the system size, hence $\nubar_m\approx 2$.

\section{Conclusions}

We have generalized an efficient algorithm to simulate the CF model with healing. This allowed to measure the critical properties of the phase transition as a function of the healing probability.
We revealed that below the critical healing probability $w_c=0.355$ the 2D interdependent network has a hybrid phase transition with both sets of exponents similar to the original model indicating universality.
\deltext{Above the critical healing however, as avalanches get eliminated the network is characterized by trivial scaling exponents in the sense that fluctuations follow the central limit theorem. The scaling relation $1-\beta_m=\gamma_a$ holds reasonably well for $w<0.1$.}
\addtext{This also means that, despite the gradual shift in the critical point due to the change in the healing probability, the critical scaling of the original CF model dominates the effects of the healing in the $w\in[0,w_c)$ range.
Above the critical healing however, the healing takes over and avalanches get eliminated. In the $w>w_c$ regime the network is characterized by trivial scaling exponents in the sense that fluctuations follow the central limit theorem.
In summary, the healing can suppress the cascades in situations where the actors of the network cannot be repaired, e.g., economic crisis situations, but the critical behavior of the phase transition does change only when the healing is higher than a threshold value.}

\begin{acknowledgments}
This work was partially supported by H2020 FETPROACT-GSS
CIMPLEX Grant No. 641191
\end{acknowledgments}

\newpage

\appendix
\section*{Appendix}


\subsection*{Algorithm}
\label{sec:healing-algo-improved}
\begin{enumerate}
\item \label{alg-heal2-attack} Let $\mathcal{D}_A$ be the set of edges to be deleted from layer $A$.
\item \label{alg-heal2-propose} Let $\mathcal{H}_A$ be the set of edges proposed as healing edges. This set is built as follows: take all the endpoints of the edges in $\mathcal{D}_A$. For each node $v$ among the endpoints list all possible pairs of the neighbors of $v$. Add the edge between each pair to $\mathcal{H}_A$ with independent probability $w$ if the edge connects two points whose dependents are in the same component in layer $B$ and the edge is not already in $\mathcal{H}_A$ nor does it exist in the network.
\item \label{alg-heal2-link} Create the edges $\mathcal{H}_A$ to the layer $A$. During the previous step, these edges were not yet added to the layer $A$ on purpose. Adding the edges in parallel with enumerating the nodes in $\mathcal{D}_A$ has unwanted side-effects that consist of nodes explored later encountering more healing links than nodes explored first. We want to avoid this and keep the algorithm independent of the order of enumeration.
\item \label{alg-heal2-remove} Remove all edges in $\mathcal{D}_A$. Whenever an edge removal splits up a connected component in $a$ into two parts, the edges that run between the parts in layer $B$ are scheduled for deletion, add them to $\mathcal{D}_B$. (This step is the analogue to immediately inactivating edges in \cite{Li2012}.)
\item \label{alg-heal2-repeat} If $\mathcal{D}_B$ is not empty, repeat the above steps swapping the roles $A\leftrightarrow B$ until no more edges are removed.
\end{enumerate}

It is clear that the link creation Step \ref{alg-heal2-link} is realized within the component before any deletion involving Step \ref{alg-heal2-remove} therefore the efficiency of the underlying dynamic graph algorithm is not degraded.


\begin{figure}[p]
\includegraphics{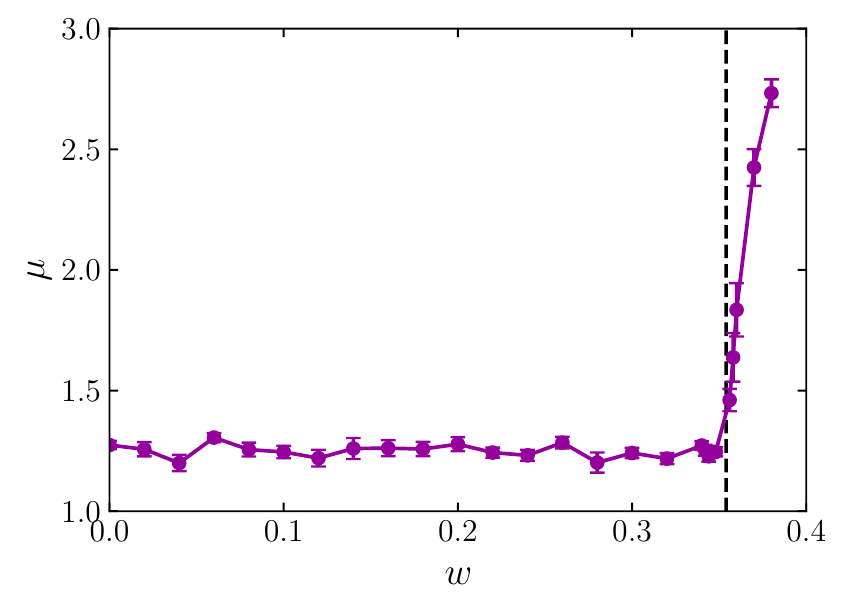}
\caption{\addtext{(color online) The execution time of our algorithm scales as $O(N^\mu)$. $\mu$ is $1.3\pm0.05$ for the small healing region $w<w_c$ and increases suddenly at the critical point $w_c$ due to the graph getting more and more connected. A complete graph of $N$ nodes would have $O(N^2)$ edges. For large $w$ each time step involves the external failure of a single node and only a limited number of cascade steps, as explained in \autoref{sec:perfect-healing}. Therefore the execution time for $w\approx1$ would scale as $O(N^3)$ plus a small correction due to the iterative cascade propagation steps but this can be verified on very small networks only.}}
\label{fig:heal-runtime-suppl}
\end{figure}

\begin{figure}[p]
\includegraphics{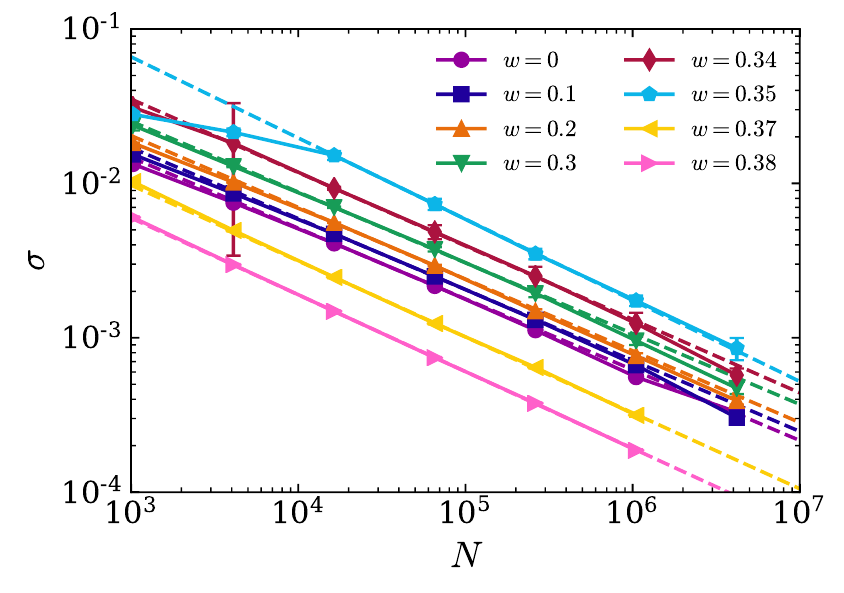}
\caption{\addtext{(color online) Examples for measuring the exponent $\nubar_m$ from $\sigma\propto N^{-1/\nubar_m}$. For $w<w_c$ we measure $\nubar_m\approx 2.2\pm0.1$ while for $w>w_c$ we argue for $\nubar_m=2$ which is confirmed by measurements.}}
\label{fig:heal-nu-m-suppl}
\end{figure}

\begin{figure}[p]
\includegraphics{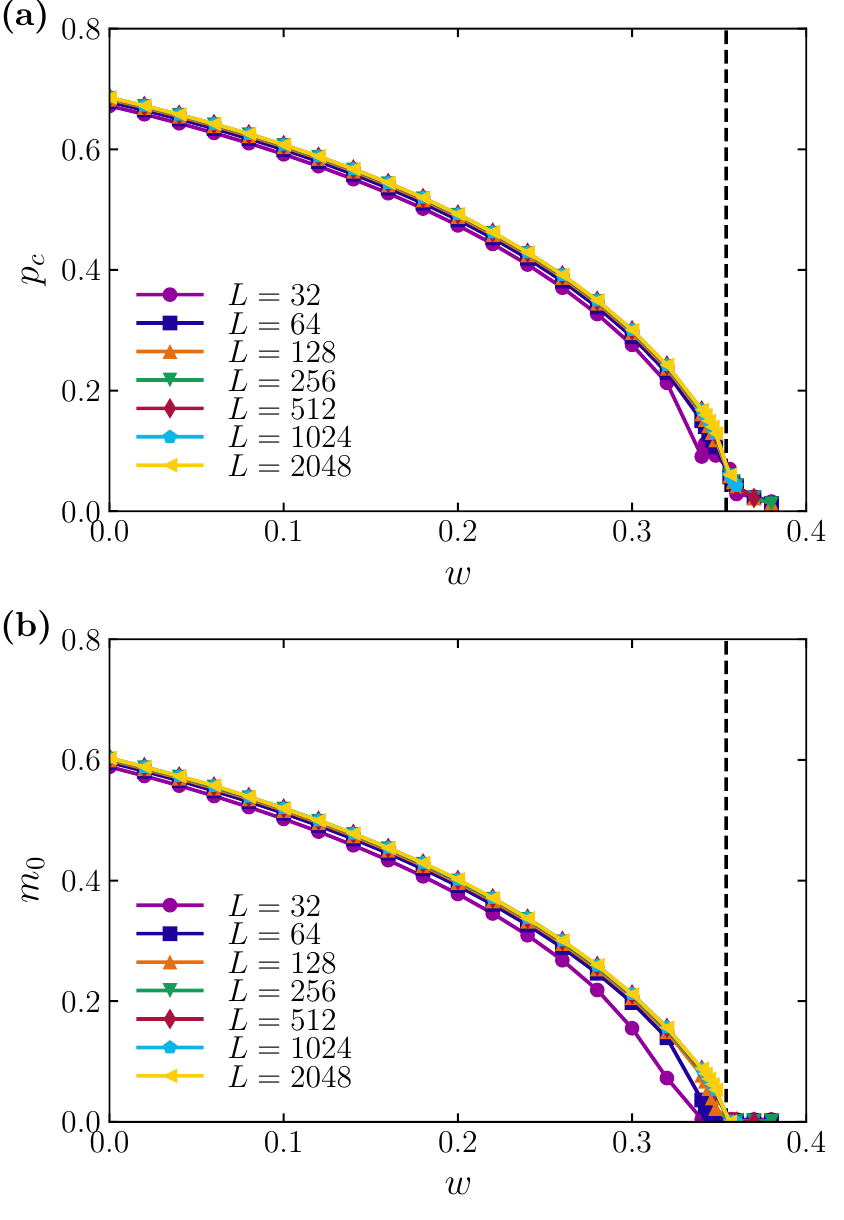}
\caption{(color online) (a) The critical point $p_c$ and (b) the related jump of the order parameter, i.e., the size of the infinite avalanche $m_0$ as a function of the system size $N=L^2$ and the healing probability $w$. Based on this information the critical values for $L=\infty$ are extrapolated using standard finite size scaling, however, near $w_c$ the extrapolation might fail. The possible failure is due to the unreliable measurement of $\nubar_m$, see \autoref{fig:heal-critical}. The finite size scaling might be further hindered by a crossover meaning that for a small fixed $w$ the order of $m_N$ is increasing with $N$ but this gets reversed at $w\approx 0.34$.}
\label{fig:heal-critical-suppl}
\end{figure}



\begin{thebibliography}{15}%
\makeatletter
\providecommand \@ifxundefined [1]{%
 \@ifx{#1\undefined}
}%
\providecommand \@ifnum [1]{%
 \ifnum #1\expandafter \@firstoftwo
 \else \expandafter \@secondoftwo
 \fi
}%
\providecommand \@ifx [1]{%
 \ifx #1\expandafter \@firstoftwo
 \else \expandafter \@secondoftwo
 \fi
}%
\providecommand \natexlab [1]{#1}%
\providecommand \enquote  [1]{``#1''}%
\providecommand \bibnamefont  [1]{#1}%
\providecommand \bibfnamefont [1]{#1}%
\providecommand \citenamefont [1]{#1}%
\providecommand \href@noop [0]{\@secondoftwo}%
\providecommand \href [0]{\begingroup \@sanitize@url \@href}%
\providecommand \@href[1]{\@@startlink{#1}\@@href}%
\providecommand \@@href[1]{\endgroup#1\@@endlink}%
\providecommand \@sanitize@url [0]{\catcode `\\12\catcode `\$12\catcode
  `\&12\catcode `\#12\catcode `\^12\catcode `\_12\catcode `\%12\relax}%
\providecommand \@@startlink[1]{}%
\providecommand \@@endlink[0]{}%
\providecommand \url  [0]{\begingroup\@sanitize@url \@url }%
\providecommand \@url [1]{\endgroup\@href {#1}{\urlprefix }}%
\providecommand \urlprefix  [0]{URL }%
\providecommand \Eprint [0]{\href }%
\providecommand \doibase [0]{http://dx.doi.org/}%
\providecommand \selectlanguage [0]{\@gobble}%
\providecommand \bibinfo  [0]{\@secondoftwo}%
\providecommand \bibfield  [0]{\@secondoftwo}%
\providecommand \translation [1]{[#1]}%
\providecommand \BibitemOpen [0]{}%
\providecommand \bibitemStop [0]{}%
\providecommand \bibitemNoStop [0]{.\EOS\space}%
\providecommand \EOS [0]{\spacefactor3000\relax}%
\providecommand \BibitemShut  [1]{\csname bibitem#1\endcsname}%
\let\auto@bib@innerbib\@empty
\bibitem [{\citenamefont {Buldyrev}\ \emph {et~al.}(2010)\citenamefont
  {Buldyrev}, \citenamefont {Parshani}, \citenamefont {Paul}, \citenamefont
  {Stanley},\ and\ \citenamefont {Havlin}}]{Buldyrev2010}%
  \BibitemOpen
  \bibfield  {author} {\bibinfo {author} {\bibfnamefont {S.~V.}\ \bibnamefont
  {Buldyrev}}, \bibinfo {author} {\bibfnamefont {R.}~\bibnamefont {Parshani}},
  \bibinfo {author} {\bibfnamefont {G.}~\bibnamefont {Paul}}, \bibinfo {author}
  {\bibfnamefont {H.~E.}\ \bibnamefont {Stanley}}, \ and\ \bibinfo {author}
  {\bibfnamefont {S.}~\bibnamefont {Havlin}},\ }\href {\doibase
  10.1038/nature08932} {\bibfield  {journal} {\bibinfo  {journal} {{Nature}}\
  }\textbf {\bibinfo {volume} {{464}}},\ \bibinfo {pages} {1025} (\bibinfo
  {year} {{2010}})}\BibitemShut {NoStop}%
\bibitem [{\citenamefont {Li}\ \emph {et~al.}(2012)\citenamefont {Li},
  \citenamefont {Bashan}, \citenamefont {Buldyrev}, \citenamefont {Stanley},\
  and\ \citenamefont {Havlin}}]{Li2012}%
  \BibitemOpen
  \bibfield  {author} {\bibinfo {author} {\bibfnamefont {W.}~\bibnamefont
  {Li}}, \bibinfo {author} {\bibfnamefont {A.}~\bibnamefont {Bashan}}, \bibinfo
  {author} {\bibfnamefont {S.~V.}\ \bibnamefont {Buldyrev}}, \bibinfo {author}
  {\bibfnamefont {H.~E.}\ \bibnamefont {Stanley}}, \ and\ \bibinfo {author}
  {\bibfnamefont {S.}~\bibnamefont {Havlin}},\ }\href {\doibase
  10.1103/PhysRevLett.108.228702} {\bibfield  {journal} {\bibinfo  {journal}
  {{Phys. Rev. Lett.}}\ }\textbf {\bibinfo {volume} {{108}}},\ \bibinfo {pages}
  {228702} (\bibinfo {year} {{2012}})}\BibitemShut {NoStop}%
\bibitem [{\citenamefont {Gao}\ \emph {et~al.}(2012{\natexlab{a}})\citenamefont
  {Gao}, \citenamefont {Buldyrev}, \citenamefont {Stanley},\ and\ \citenamefont
  {Havlin}}]{Havlin2012}%
  \BibitemOpen
  \bibfield  {author} {\bibinfo {author} {\bibfnamefont {J.}~\bibnamefont
  {Gao}}, \bibinfo {author} {\bibfnamefont {S.~V.}\ \bibnamefont {Buldyrev}},
  \bibinfo {author} {\bibfnamefont {H.~E.}\ \bibnamefont {Stanley}}, \ and\
  \bibinfo {author} {\bibfnamefont {S.}~\bibnamefont {Havlin}},\ }\href
  {\doibase 10.1038/nphys2180} {\bibfield  {journal} {\bibinfo  {journal}
  {{Nat. Phys.}}\ }\textbf {\bibinfo {volume} {{8}}},\ \bibinfo {pages} {40}
  (\bibinfo {year} {{2012}}{\natexlab{a}})}\BibitemShut {NoStop}%
\bibitem [{\citenamefont {Gao}\ \emph {et~al.}(2012{\natexlab{b}})\citenamefont
  {Gao}, \citenamefont {Buldyrev}, \citenamefont {Havlin},\ and\ \citenamefont
  {Stanley}}]{Buldyrev2012}%
  \BibitemOpen
  \bibfield  {author} {\bibinfo {author} {\bibfnamefont {J.}~\bibnamefont
  {Gao}}, \bibinfo {author} {\bibfnamefont {S.~V.}\ \bibnamefont {Buldyrev}},
  \bibinfo {author} {\bibfnamefont {S.}~\bibnamefont {Havlin}}, \ and\ \bibinfo
  {author} {\bibfnamefont {H.~E.}\ \bibnamefont {Stanley}},\ }\href {\doibase
  10.1103/PhysRevE.85.066134} {\bibfield  {journal} {\bibinfo  {journal}
  {{Phys. Rev. E}}\ }\textbf {\bibinfo {volume} {{85}}},\ \bibinfo {pages}
  {066134} (\bibinfo {year} {{2012}}{\natexlab{b}})}\BibitemShut {NoStop}%
\bibitem [{\citenamefont {Zhou}\ \emph {et~al.}(2013)\citenamefont {Zhou},
  \citenamefont {Gao}, \citenamefont {Stanley},\ and\ \citenamefont
  {Havlin}}]{Zhou2013}%
  \BibitemOpen
  \bibfield  {author} {\bibinfo {author} {\bibfnamefont {D.}~\bibnamefont
  {Zhou}}, \bibinfo {author} {\bibfnamefont {J.}~\bibnamefont {Gao}}, \bibinfo
  {author} {\bibfnamefont {H.~E.}\ \bibnamefont {Stanley}}, \ and\ \bibinfo
  {author} {\bibfnamefont {S.}~\bibnamefont {Havlin}},\ }\href {\doibase
  10.1103/PhysRevE.87.052812} {\bibfield  {journal} {\bibinfo  {journal} {Phys.
  Rev. E}\ }\textbf {\bibinfo {volume} {87}},\ \bibinfo {pages} {052812}
  (\bibinfo {year} {2013})},\ \Eprint {http://arxiv.org/abs/arXiv:1206.2427v2}
  {arXiv:arXiv:1206.2427v2} \BibitemShut {NoStop}%
\bibitem [{\citenamefont {Majdandzic}\ \emph {et~al.}(2016)\citenamefont
  {Majdandzic}, \citenamefont {Braunstein}, \citenamefont {Curme},
  \citenamefont {Vodenska}, \citenamefont {Levy-Carciente}, \citenamefont
  {{Eugene Stanley}},\ and\ \citenamefont {Havlin}}]{Majdandzic2016}%
  \BibitemOpen
  \bibfield  {author} {\bibinfo {author} {\bibfnamefont {A.}~\bibnamefont
  {Majdandzic}}, \bibinfo {author} {\bibfnamefont {L.~A.}\ \bibnamefont
  {Braunstein}}, \bibinfo {author} {\bibfnamefont {C.}~\bibnamefont {Curme}},
  \bibinfo {author} {\bibfnamefont {I.}~\bibnamefont {Vodenska}}, \bibinfo
  {author} {\bibfnamefont {S.}~\bibnamefont {Levy-Carciente}}, \bibinfo
  {author} {\bibfnamefont {H.}~\bibnamefont {{Eugene Stanley}}}, \ and\
  \bibinfo {author} {\bibfnamefont {S.}~\bibnamefont {Havlin}},\ }\href
  {\doibase 10.1038/ncomms10850} {\bibfield  {journal} {\bibinfo  {journal}
  {Nature Communications}\ }\textbf {\bibinfo {volume} {7}},\ \bibinfo {pages}
  {10850} (\bibinfo {year} {2016})},\ \Eprint {http://arxiv.org/abs/1502.00244}
  {arXiv:1502.00244} \BibitemShut {NoStop}%
\bibitem [{\citenamefont {{Di Muro}}\ \emph {et~al.}(2016)\citenamefont {{Di
  Muro}}, \citenamefont {{La Rocca}}, \citenamefont {Stanley}, \citenamefont
  {Havlin},\ and\ \citenamefont {Braunstein}}]{DiMuro2016}%
  \BibitemOpen
  \bibfield  {author} {\bibinfo {author} {\bibfnamefont {M.~A.}\ \bibnamefont
  {{Di Muro}}}, \bibinfo {author} {\bibfnamefont {C.~E.}\ \bibnamefont {{La
  Rocca}}}, \bibinfo {author} {\bibfnamefont {H.~E.}\ \bibnamefont {Stanley}},
  \bibinfo {author} {\bibfnamefont {S.}~\bibnamefont {Havlin}}, \ and\ \bibinfo
  {author} {\bibfnamefont {L.~A.}\ \bibnamefont {Braunstein}},\ }\href
  {\doibase 10.1038/srep22834} {\bibfield  {journal} {\bibinfo  {journal} {Sci.
  Rep.}\ }\textbf {\bibinfo {volume} {6}},\ \bibinfo {pages} {22834} (\bibinfo
  {year} {2016})},\ \Eprint {http://arxiv.org/abs/1512.02555}
  {arXiv:1512.02555} \BibitemShut {NoStop}%
\bibitem [{\citenamefont {Stippinger}\ and\ \citenamefont
  {Kert\'esz}(2014)}]{Stippinger2014}%
  \BibitemOpen
  \bibfield  {author} {\bibinfo {author} {\bibfnamefont {M.}~\bibnamefont
  {Stippinger}}\ and\ \bibinfo {author} {\bibfnamefont {J.}~\bibnamefont
  {Kert\'esz}},\ }\href {\doibase 10.1016/j.physa.2014.08.069} {\bibfield
  {journal} {\bibinfo  {journal} {Physica A: Statistical Mechanics and its
  Applications}\ }\textbf {\bibinfo {volume} {416}},\ \bibinfo {pages} {481}
  (\bibinfo {year} {2014})},\ \Eprint {http://arxiv.org/abs/1312.1993}
  {arXiv:1312.1993} \BibitemShut {NoStop}%
\bibitem [{\citenamefont {Lee}\ \emph {et~al.}(2016)\citenamefont {Lee},
  \citenamefont {Choi}, \citenamefont {Stippinger}, \citenamefont {Kert\'esz},\
  and\ \citenamefont {Kahng}}]{Lee2016}%
  \BibitemOpen
  \bibfield  {author} {\bibinfo {author} {\bibfnamefont {D.}~\bibnamefont
  {Lee}}, \bibinfo {author} {\bibfnamefont {S.}~\bibnamefont {Choi}}, \bibinfo
  {author} {\bibfnamefont {M.}~\bibnamefont {Stippinger}}, \bibinfo {author}
  {\bibfnamefont {J.}~\bibnamefont {Kert\'esz}}, \ and\ \bibinfo {author}
  {\bibfnamefont {B.}~\bibnamefont {Kahng}},\ }\href {\doibase
  10.1103/PhysRevE.93.042109} {\bibfield  {journal} {\bibinfo  {journal}
  {{Phys. Rev. E}}\ }\textbf {\bibinfo {volume} {93}},\ \bibinfo {pages}
  {042109} (\bibinfo {year} {2016})}\BibitemShut {NoStop}%
\bibitem [{\citenamefont {Hwang}\ \emph {et~al.}(2015)\citenamefont {Hwang},
  \citenamefont {Choi}, \citenamefont {Lee},\ and\ \citenamefont
  {Kahng}}]{Hwang2015}%
  \BibitemOpen
  \bibfield  {author} {\bibinfo {author} {\bibfnamefont {S.}~\bibnamefont
  {Hwang}}, \bibinfo {author} {\bibfnamefont {S.}~\bibnamefont {Choi}},
  \bibinfo {author} {\bibfnamefont {D.}~\bibnamefont {Lee}}, \ and\ \bibinfo
  {author} {\bibfnamefont {B.}~\bibnamefont {Kahng}},\ }\href {\doibase
  10.1103/PhysRevE.91.022814} {\bibfield  {journal} {\bibinfo  {journal}
  {{Phys. Rev. E}}\ }\textbf {\bibinfo {volume} {91}},\ \bibinfo {pages}
  {022814} (\bibinfo {year} {2015})}\BibitemShut {NoStop}%
\bibitem [{\citenamefont {Zhou}\ \emph {et~al.}(2014)\citenamefont {Zhou},
  \citenamefont {Bashan}, \citenamefont {Cohen}, \citenamefont {Berezin},
  \citenamefont {Shnerb},\ and\ \citenamefont {Havlin}}]{Zhou2014}%
  \BibitemOpen
  \bibfield  {author} {\bibinfo {author} {\bibfnamefont {D.}~\bibnamefont
  {Zhou}}, \bibinfo {author} {\bibfnamefont {A.}~\bibnamefont {Bashan}},
  \bibinfo {author} {\bibfnamefont {R.}~\bibnamefont {Cohen}}, \bibinfo
  {author} {\bibfnamefont {Y.}~\bibnamefont {Berezin}}, \bibinfo {author}
  {\bibfnamefont {N.}~\bibnamefont {Shnerb}}, \ and\ \bibinfo {author}
  {\bibfnamefont {S.}~\bibnamefont {Havlin}},\ }\href {\doibase
  10.1103/PhysRevE.90.012803} {\bibfield  {journal} {\bibinfo  {journal} {Phys.
  Rev. E}\ }\textbf {\bibinfo {volume} {90}},\ \bibinfo {pages} {012803}
  (\bibinfo {year} {2014})},\ \Eprint {http://arxiv.org/abs/1211.2330}
  {arXiv:1211.2330} \BibitemShut {NoStop}%
\bibitem [{\citenamefont {Grassberger}(2015)}]{Grassberger2015}%
  \BibitemOpen
  \bibfield  {author} {\bibinfo {author} {\bibfnamefont {P.}~\bibnamefont
  {Grassberger}},\ }\href {\doibase {10.1103/PhysRevE.91.062806}} {\bibfield
  {journal} {\bibinfo  {journal} {{Phys. Rev. E}}\ }\textbf {\bibinfo {volume}
  {{91}}},\ \bibinfo {pages} {062806} (\bibinfo {year} {{2015}})}\BibitemShut
  {NoStop}%
\bibitem [{\citenamefont {Holm}\ \emph {et~al.}(2001)\citenamefont {Holm},
  \citenamefont {de~Lichtenberg},\ and\ \citenamefont {Thorup}}]{Holm2001}%
  \BibitemOpen
  \bibfield  {author} {\bibinfo {author} {\bibfnamefont {J.}~\bibnamefont
  {Holm}}, \bibinfo {author} {\bibfnamefont {K.}~\bibnamefont
  {de~Lichtenberg}}, \ and\ \bibinfo {author} {\bibfnamefont {M.}~\bibnamefont
  {Thorup}},\ }\href {\doibase 10.1145/502090.502095} {\bibfield  {journal}
  {\bibinfo  {journal} {J. ACM}\ }\textbf {\bibinfo {volume} {48}},\ \bibinfo
  {pages} {723} (\bibinfo {year} {2001})}\BibitemShut {NoStop}%
\bibitem [{Note1()}]{Note1}%
  \BibitemOpen
  \bibinfo {note} {{\protect \addtext{This quantity is different from
  $1/\alpha $ introduced in \cite {Zhou2014} as explained in the following. In
  the case the avalanche is triggered by an initial removal of a fraction of
  the nodes the exponent $1/\alpha $ describes the finite size scaling of the
  avalanche duration (the number of back-and-forth propagation steps). However,
  in the healing model the avalanches are triggered by repeated single node
  removals therefore the avalanche dynamics is different. In this case a
  similar quantity, the exponent $\protect \mathaccentV {bar}016{\nu }_a$ was
  introduced but they may eventually differ in value.}}}\BibitemShut {Stop}%
\bibitem [{Note2()}]{Note2}%
  \BibitemOpen
  \bibinfo {note} {Unfortunately one cannot gain insight by experimenting with
  $p_c$ to get satisfactory scaling because $m_0$ is to be determined in
  parallel making the experimentation prone to errors.}\BibitemShut {Stop}%
\end{thebibliography}

%

\end{document}